\documentclass[12pt]{article}
\usepackage{amsmath,amssymb,amsfonts,hyperref,dsfont}
\usepackage{blindtext}
\usepackage[english]{babel}
\usepackage{graphicx}
\usepackage{subcaption} 
\usepackage[utf8]{inputenc}
\usepackage[square]{natbib}
\setcitestyle{numbers}

\allowdisplaybreaks
\title{Parametric and non-parametric for extreme earthquake events \\
 the joint tail inference for mainshocks and aftershocks}

\def\P{\mathbb{P}}
\def\E{\mathbb{E}}

\title{Parametric and non-parametric estimation of extreme earthquake event: the joint tail inference for mainshocks and aftershocks}

\author{Juan-Juan Cai\thanks{Department of Applied Mathematics, Delft University of Technology, Mekelweg 4 2628 CD Delft, the Netherlands; email: j.j.cai@tudelft.nl}
\and  Phyllis Wan\thanks{Econometric Institute, Erasmus University Rotterdam, Burg.\ Oudlaan 50, 3062 PA Rotterdam, the Netherlands; email: wan@ese.eur.nl}
\and Gamze Ozel\thanks{Department of Statistics, Hacettepe University, 06800 Ankara, Turkey; gamzeozl@hacettepe.edu.tr}
}

\begin{document}

\maketitle

\begin{abstract}

In an earthquake event, the combination of a strong mainshock and damaging aftershocks is often the cause of severe structural damages and/or high death tolls.  The objective of this paper is to provide estimation for the probability of such extreme events where the mainshock and the largest aftershocks exceed certain thresholds.  Two approaches are illustrated and compared -- a parametric approach based on previously observed stochastic laws in earthquake data, and a non-parametric approach based on bivariate extreme value theory.  We analyze the earthquake data from the North Anatolian Fault Zone (NAFZ) in Turkey during 1965--2018 and show that the two approaches provide unifying results.

\end{abstract}


\section{Introduction}\label{Sec:Intro}

In a seismically active area, a strong earthquake, namely the {\em mainshock}, is often followed by subsequent damaging earthquakes, known as the {\em aftershocks}.  These aftershocks may occur in numerous quantity and with magnitudes equivalent to powerful earthquakes on their own.  For instance, in the 1999 \.{I}zmit earthquake, a magnitude 7.6 mainshock triggered hundreds of aftershocks with magnitudes greater than or equal to 4 in the first six days, cf.~\cite{Polatetal2002}.  In the 2008 Sichuan earthquake, a mainshock of magnitude 8.0 induced a series of aftershocks with magnitudes up to 6.0.  The results are severe structural damage and loss of life, especially when the area has already been weakened by the mainshock.  The \.{I}zmit earthquake killed over 17,000 people and left half a million homeless \cite{marza2004}.  The Sichuan earthquake caused over 69,000 deaths and damages of over 150 billion US dollars \cite{cui2011}. 

The goal of this paper is to provide a statistical analysis for the joint event of an extreme mainshock and extreme aftershocks. Throughout the paper, we denote the magnitude of a mainshock with $X$ and that of the largest aftershock with $Y$. We estimate via two approaches the probability of
\begin{equation}
\P(X>s, Y>t),  \label{eq:pst}
\end{equation}
for large values of $s$ and $t$.
The first approach uses a parametric model based on a series of well-know stochastic laws that describe the empirical relationships of the aftershocks and the mainshock, which we briefly review in Section~\ref{subsec:par}.  In the second approach, we apply bivariate extreme value theory to estimate the joint tail.  Both methods are applied to the extreme earthquake events in the North Anatolian Fault Zone (NAFZ) in Turkey, the region where the 1999 \.{I}zmit earthquake occurred.

The remainder of the paper is structured as follows.  In Section~\ref{Sec:Data}, we present the earthquake data in NAFZ and describe the relevant data processing.  Section~\ref{Sec:Method} provides the parametric and non-parametric estimation procedures for the joint main-/after-shock distribution.  The detailed data analysis and results are presented in Section~\ref{Sec:Result}.  We conclude in Section~\ref{Sec:Discuss} with some discussions.




\section{Data description}\label{Sec:Data}

We use the North Anatolian Fault Zone (NAFZ) as an area of investigation due to its long and extensive historical record of large earthquakes \cite{Ambraseys1970,AmbraseysFinkel1987}.  Extending from eastern Turkey to Greece, the 1,500-kilometer-long rip sustained several cycle-like sequences of large-magnitude ($M>7$) earthquakes over the past centuries \cite{Steinetal1997}, several resulting in high death tolls and severe economic losses.  The most recent activities include the \.{I}zmit (Mw 7.6) and D\"uzce (Mw 7.1) earthquakes of 1999 \cite{Parsonsetal2000,Reilingeretal2000}.\footnote{Regarding the earthquake scale in our data: Before 1977, all earthquakes were recorded in the body-wave magnitude scale (mb) or the surface-wave magnitude scale (Ms), depending on the depth of the earthquake.  Following the development of the moment magnitude scale (Mw) by \cite{kanamori1977,hanks1979}, earthquakes with magnitude larger than 5.0 are recorded in the Mw scale whereas smaller earthquakes were still generally measure in the mb or the Ms scale.  From 2012 on, all earthquakes are recorded in the Mw scale.}  

\begin{figure}[ht]
\centering
\begin{subfigure}{.49\textwidth}
\centering
\includegraphics[width=\linewidth]{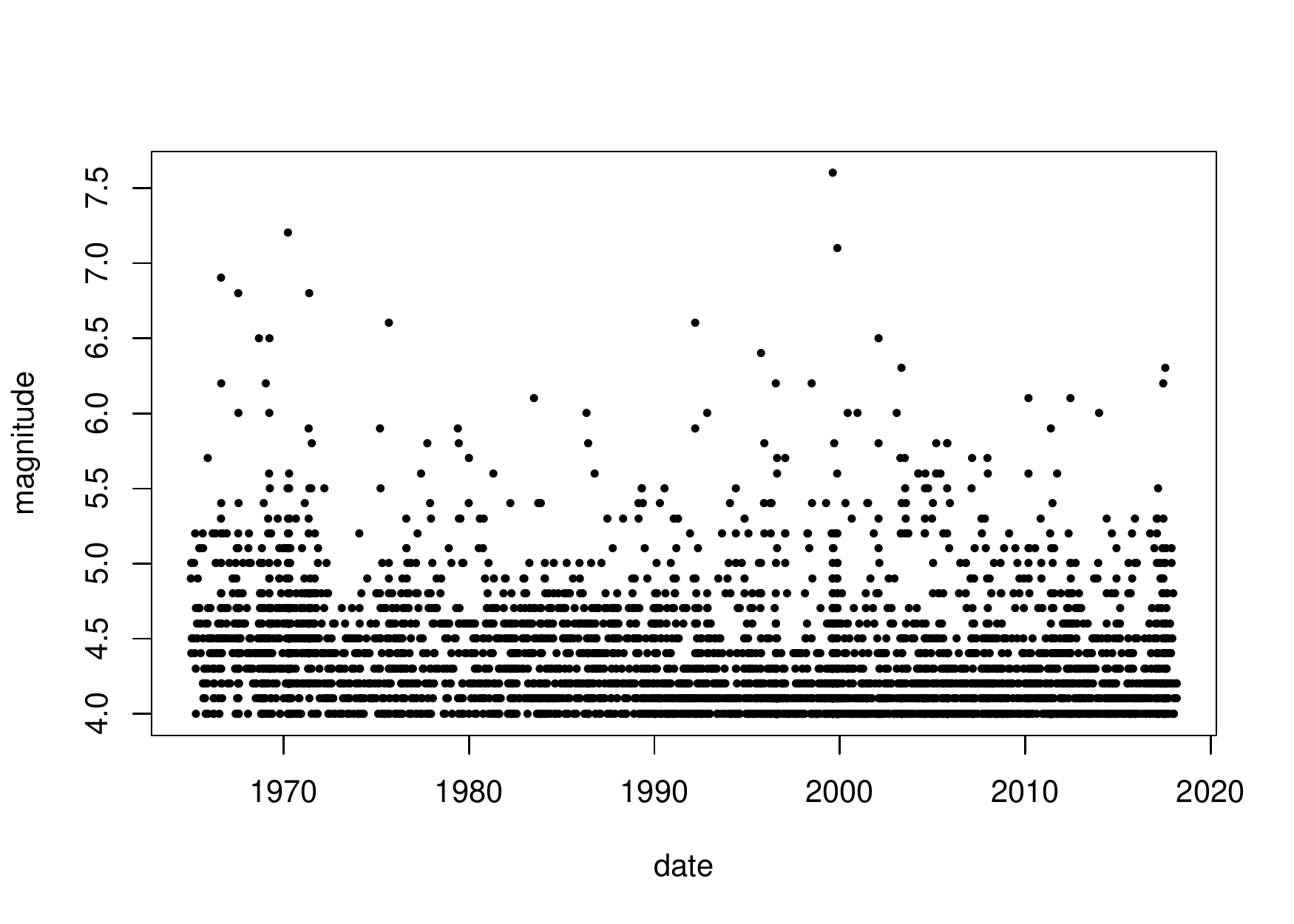}
\end{subfigure}
\begin{subfigure}{.49\textwidth}
\centering
\includegraphics[width=\linewidth]{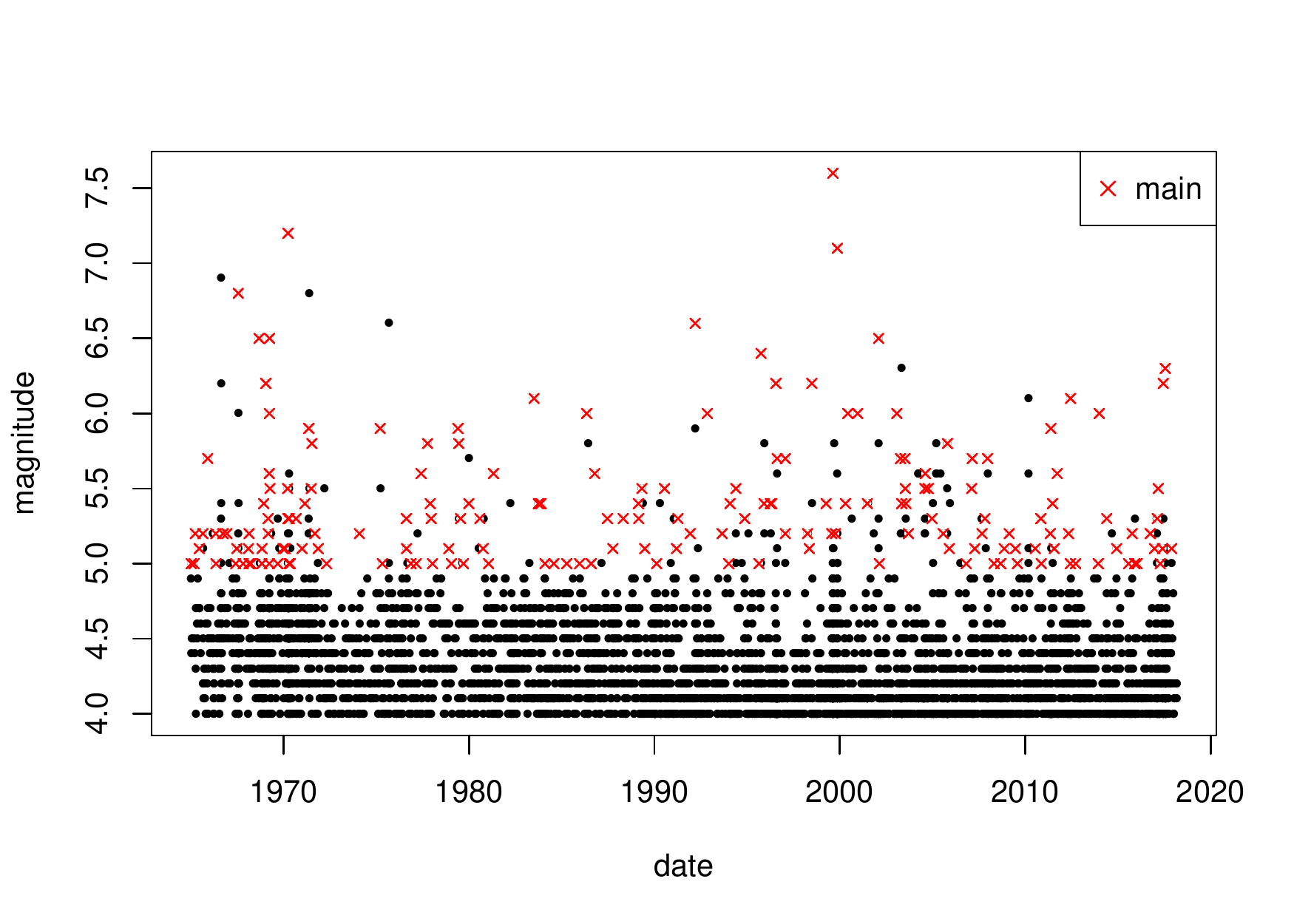}
\end{subfigure}
\caption{Shocks and labelled mainshocks in the NAFZ during 1965--2018.}
\label{fig:shocks}
\end{figure}

We obtain data from the Presidential of Earthquake Department database of the Turkish Disaster and Emergency Management Authority (\url{https://deprem.afad.gov.tr/?lang=en}) and consider all earthquake records between 1965--2018 with magnitudes 4 or higher in the area of $39.00^o - 42.00^o$ latitude and $26.00^o-40.00^o$ longitude.  The left panel of Figure~\ref{fig:shocks} shows the time series of all earthquakes from 1965 onward.

We now label earthquake events by identifying the mainshocks and their corresponding aftershocks.  Being interested in extreme events, we only consider earthquake events with significant mainshocks such that $X\ge5$.  We use the window algorithm proposed in \cite{gardner1974} as follows.  For each shock with magnitude $X\ge5$, we scan the window within distance $L(X)$ and time $T(X)$.  If a larger shock exists, we move on to that shock and perform the same scan.  If not, then the shock is labelled as the mainshock and all shocks within the specified window are pronounced as its aftershocks.  Table~\ref{tab:aftershock} provides the values for $L(X)$ and $T(X)$.  For example, for an earthquake of magnitude 6.0,  any shock  following $T=510$ days and within $L=54$ km radius, with a magnitude less than 6, is considered to be its aftershock.  

The right panel of Figure~\ref{fig:shocks} shows the labelled mainshocks in the time series.  The algorithm identifies $n=180$ earthquake events with mainshocks $X\ge5$ among which 129 have aftershocks with magnitude greater than 4.  Note that a few large earthquakes in the early years are not labelled as mainshocks, instead they are identified as aftershocks of mainshocks before the year 1965 from earlier data.

\begin{table}
\begin{center}
\begin{tabular}{ccc}
\hline
$X$ & $L$ & $T$ \\
& (km) & (days)\\
\hline
5.0--5.4 & 40 & 155 \\
5.5--5.9 & 47 & 290 \\
6.0--6.4 & 54 & 510 \\
6.5--6.9 & 61 & 790 \\
7.0--7.4 & 70 & 915 \\
7.5--7.9 & 81 & 960 \\
8.0--8.4 & 94 & 985 \\
\hline
\end{tabular}
\caption{Window specification $L(X)$, $T(X)$ for aftershock labelling from \cite{gardner1974}.}
\label{tab:aftershock}
\end{center}
\end{table}

%
%


\section{Methodology}\label{Sec:Method}

\subsection{Parametric approach} \label{subsec:par}

It is agreed in the literature that the distribution of aftershocks in space, time and magnitude can be characterized by stochastic laws, see \cite{Utsu1970, Utsu1971} and \cite{Utsu1972} for a summary with detailed empirical studies. In this section, we propose a simple parametric model for the joint magnitudes of the mainshock and the largest aftershock based on these relationships.  This derivation is similar to that in \cite{Vere2006}.

The following empirical evidence for aftershocks have been noted in prior literature.
\begin{enumerate}
\item
	The frequency $g(t)$ of the aftershocks per time unit at time $t$ after the mainshock follows the modified Omori's law: 
	$$
		g(t) = \frac{K}{(t+c)^p},
	$$
	where $K, c, p$ are constants \cite{Utsu1970}.
\item
	The magnitude of the aftershocks follows Gutenberg-Richter's law \cite{GutenbergRichter1944}, that is, the number of aftershocks $N(m)$ with magnitude $m$ follows
	\begin{equation} \label{eq:gr}
		N(m) = 10^{a-bm},
	\end{equation}
	where $a,b$ are constants.
\item
	The magnitude difference between a mainshock and its largest aftershock is approximately constant, independent of the mainshock magnitude and typically between 1.1 and 1.2 \cite{Bath1965}.
\end{enumerate}
Based on the above, \cite{Utsu1970} modelled the intensity rate of aftershocks with magnitude $m$ as 
\begin{equation} \label{eq:as_intensity}
	\lambda(t,m) = \frac{10^{a+b(m_0-m)}}{(t+c)^p}, \quad m \le m_0,
\end{equation}
where $m_0$ is the mainshock magnitude and $a,b,c,p$ are constants.   By defintion, $m\le m_0$ such that aftershocks are always smaller than the mainshock.  This modelling is used widely in ensuing literature, cf.~\cite{ReasenbergJones1989}, and is the basis of the ETAS (epidemic-type aftershock sequence) simulation model, cf.~\cite{Ogata1988}.  It is common to model the occurrences of aftershocks as a Poisson point process.

On the other hand, the mainshocks can be considered as independent events and their magnitude can also be modelled by the Gutenberg-Richter's law in equation \eqref{eq:gr} \cite{Utsu1972}.  In the following, we model the magnitude of the mainshocks $X$ using an exponential distribution with distribution and density functions
\begin{equation}\label{eq:ms}
	\P(X>x) = e^{-\alpha x}, \quad f_X(x) = \alpha e^{-\alpha x}.
\end{equation}

\subsubsection{The model}

Let  $X_A$ denote the magnitude of an aftershock.  Given the mainshock $X=m_0$, we assume that the aftershocks sequence follows a non-homogeneous Poisson process with intensity function \eqref{eq:as_intensity}.  We derive the following.
\begin{itemize}
\item
	The total number of aftershocks $N$ follows a Poisson random variable with mean
$$
	\E[N|X=m_0] = \sum_{t=1}^\infty \int_0^{m_0} \lambda(t,u) du =: Ce^{\beta m_0} \left(1-e^{-\beta m_0}\right),
$$
where $\beta = b \ln 10$ and $C = \frac{1}\beta10^a\sum_{t=1}^\infty \frac{1}{(t+c)^p}$.
\smallskip
\item
	The conditional distribution of $X_A$ follows
$$
	\P(X_A>m|X=m_0) = \frac{\sum_{t=1}^\infty \int_m^{m_0} \lambda(t,u) du}{\sum_{t=1}^\infty \int_0^{m_0} \lambda(t,u) du}  = \frac{e^{-\beta m}-e^{-\beta m_0}}{1-e^{-\beta m_0}}, \quad 0 \le m \le m_0,
$$
and is conditionally independent of $N$.
\end{itemize}
Observe that the largest aftershock $Y=\max_{1\leq i\leq N}X_{A_i}$. Therefore, by the conditional independence of $N$ and $X_A$, it follows that for $m\in [0, m_0]$,
	\begin{eqnarray*}
		\P(Y\le m|X=m_0) &=& \E\left[(1-P(X_A>m))^N|X=m_0\right] \\
		&=& \exp\left\{-\P(X_A>m|X=m_0){\E[N|X=m_0]}\right\} \\
		&=& \exp\left\{-\frac{e^{-\beta m}-e^{-\beta m_0}}{1-e^{-\beta m_0}} \cdot Ce^{\beta m_0} \left(1-e^{-\beta m_0}\right) \right\} \\
		&=& \exp\left\{-C\left(e^{-\beta (m-m_0)}-1\right)\right\}, 
	\end{eqnarray*}
where the second equality follows from a property of Poisson expectation: $\E[(1-p)^N]=\exp(-p\lambda)$, where $N\sim \text{Poi}(\lambda)$ and $p \in (0, 1)$. 
Let $Z := X - Y$, then $Z$ has distribution function
\begin{equation} \label{eq:gompertz}
	F(z) := P(Z\ge z) = \exp\left\{-C\left(e^{\beta z}-1\right)\right\}, \quad z\le m_0.
\end{equation}
This suggests that $Z$ follows a Gompertz distribution, that is, $-Z$ follows a Gumbel distribution conditional to be negative. If we impose the convention that $\{Z>m_0\}=\{X_A<0\}$ represents the event that no aftershock occurs, then we can model $Z$ as independent of $X$. Note that when $m_0$ is large the probability of $\{Z>m_0\}$ is negligible.

Combining \eqref{eq:ms} and \eqref{eq:gompertz} yields the   joint model for $(X,Y)$  given by
$$
\P(X>x,Y>y) = \P(X>x,Z<X-y) = \int_x^\infty f_X(x) \int_0^{x-y} f_Z(z) dz dx,
$$
where $f_X(x)$ is as defined in \eqref{eq:ms} and $f_Z(z)$ is the density function of $Z$ from \eqref{eq:gompertz}.  Given data, the parameters $(\alpha,\beta,C)$ can be estimated through maximum likelihood.


\subsection{Bivariate extreme value approach}
Multivariate extreme statistics has been exhibited to be a powerful tool for inference on multidimensional risk factors.  Examples of applications can be found in \cite{dehaan1998, ledford1997} and \cite{PoonRockingerTawn2004} among others. Recall that the goal is to estimate the probability: $\P(X>t, Y>s)$. To this end, we assume that the joint distribution of $(X,Y)$ is in the max domain of a bivariate extreme distribution introduced in \cite{deHaanResnick1977}.  This is a common condition in multivariate tail analysis and includes distributions with various types of copulas.
 Let $F_1$ and $F_2$ denote the marginal distribution functions of $X$ and $Y$, respectively. The assumption implies that for any $(x, y)\in [0, \infty]^2 \setminus ({\infty, \infty})$, the following limit exists:
\begin{equation}
\lim_{t\rightarrow 0}\frac{1}{t}\P(1-F_1(X)<tx, 1-F_2(y)<ty)=:R(x,y).  \label{eq: R}
\end{equation}
The function $R$ characterizes the extremal dependence between $X$ and $Y$ and it can be expressed via other extremal dependence measures. For instance, it is linked to the stable tail dependence function $L$ and the Pickand function $A$:
\begin{equation}
R(x,y)=x+y-L(x, y)=(x+y)\left(1-A\left(\frac{y}{x+y}\right)\right). \label{eq: RLA}
\end{equation}
 For a general review on the multivariate extreme value theory, see for example Chapter 6 in \cite{deHaanFerreira2006} and Chapter 8 in \cite{Beirlantetal2004}.

The limit relation in \eqref{eq: R} guarantees the regularity in the right tail of the copula of $(X, Y)$, which enables us to do the bivariate extrapolation to the range far beyond the historical observations. 
Let $s$ and $t$ be sufficiently large and denote that $p_1=\P(X>s)$ and $p_2=\P(X>t)$.
\begin{eqnarray}
\P(X>s, Y>t)&=&\P(1-F_1(X)<p_1, 1-F_2(y)<p_2)\nonumber\\
&=&p_2\cdot \frac{1}{p_2}\P\left(1-F_1(X)<p_2\cdot \frac{p_1}{p_2}, 1-F_2(y)<p_2\right) \nonumber\\
&\approx &p_2 R\left(\frac{p_1}{p_2},1\right). \label{eq:pst}
\end{eqnarray}
Then the problem transforms to estimating $p_1$, $p_2$ and $R(x, 1)$. Due to the relation in \eqref{eq: RLA}, the various methods of estimating $L$ or $A$ can be applied to estimate $R(x, 1)$; for instance see \cite{Caperaaetal1997, Einmahletal2008,  Bucheretal2011,Fougeresetal2015, Beirlantetal2016} among many others. 
Because of the particular features of earthquake data -- that they have been rounded to the first digit and censored below -- we use a basic non-parametric estimator of $R(x, 1)$, which
requires least assumptions on the data and is the basis of other more advanced estimation approaches. Let $n$ be the sample size and $k=k(n)$ be a sequence of integers such that $k\rightarrow\infty$ and $k/n\rightarrow 0$ as $n\rightarrow \infty$.
Let $R_i^X$ and $R_i^Y$ denote the ranks of $X_i$ and $Y_i$ in their respective samples. The estimator of $R(x,1)$ is given by
\begin{align}
\hat R(x, 1)=\frac{1}{k}\sum_{i=1}^n I(R_i^X>n+1/2-kx, R_i^Y>n+1/2-k). \label{eq: hatR}
\end{align}

As for estimating $p_1$ and $p_2$, we fit exponential distributions to both margins, which is a typical choice for modeling earthquake magnitude justified by the Gutenberg–Richter law \cite{GutenbergRichter1944}. A natural alternative is to apply univariate extrem value theory to estimate these tail probabilities. Many studies have  been devoted to study the tail distribution or the endpoint of earthquake magnitude; see for instance \cite{Kijko2004} and \cite{Beirlantetal2018}. However, due to the small sample size and the rounding issue, we choose to fit  parametric margins.  


\section{Results}\label{Sec:Result}

\begin{figure}[ht]
\centering
\begin{subfigure}{.49\textwidth}
\centering
\includegraphics[width=\linewidth]{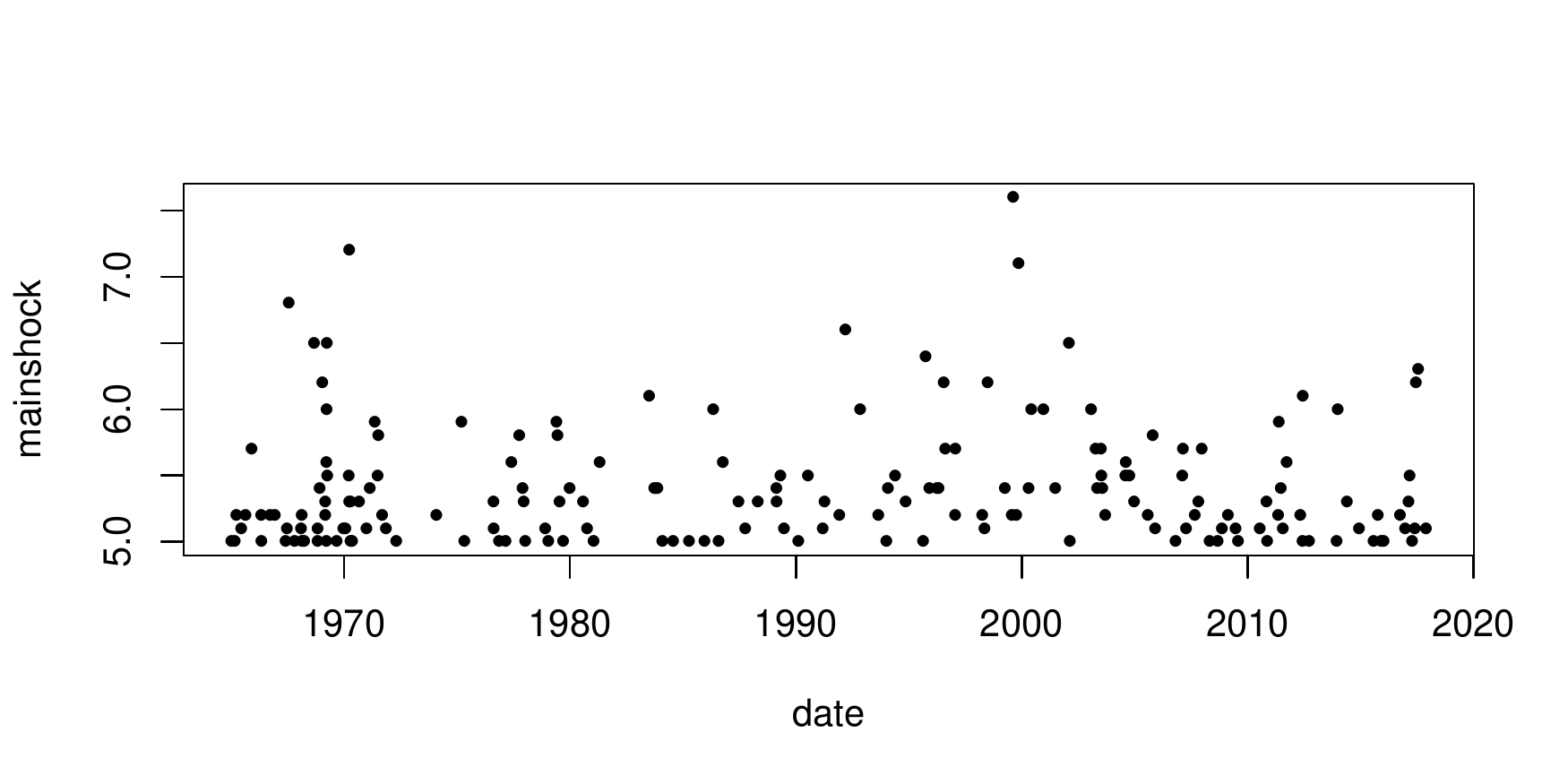}
\end{subfigure}
\begin{subfigure}{.49\textwidth}
\centering
\includegraphics[width=\linewidth]{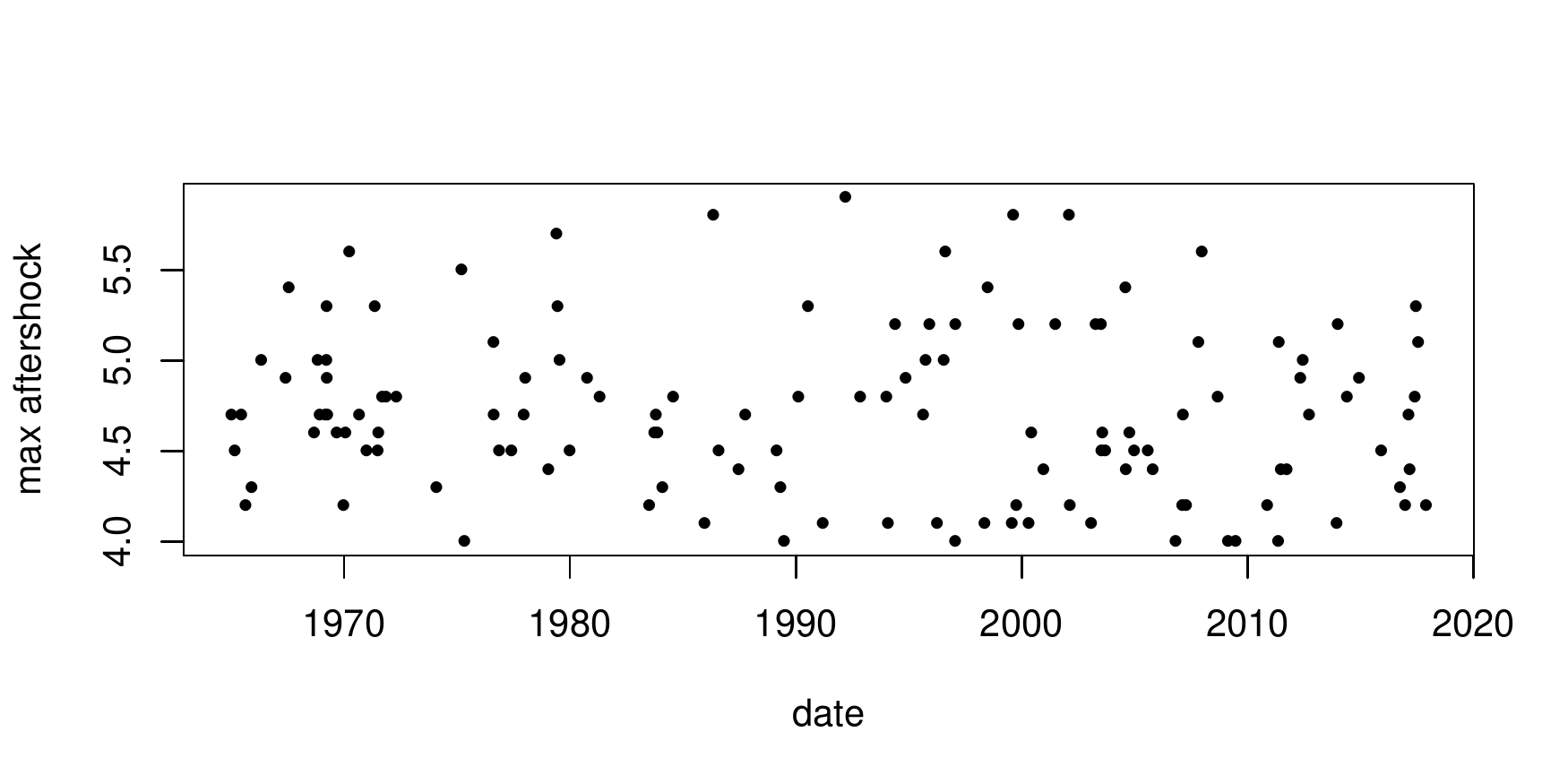}
\end{subfigure}
\caption{Time series of mainshocks (left) and largest aftershocks (right).}
\label{fig:ts}
\end{figure}

From Section~\ref{Sec:Data} we extract from the NAFZ dataset time series of mainshocks magnitude $(x_i)$ where $x_i\ge5$.  For the time series of the corresponding largest aftershock, we only observe the values that are above 4, that is, we observe $\left(y_i \bf1_{\{y_i\ge4\}}\right)$.  The two time series are plotted in Figure~\ref{fig:ts}. 

\subsection{Parametric approach}

\begin{figure}[ht]
\centering
\begin{subfigure}{.49\textwidth}
\centering
\includegraphics[width=\linewidth]{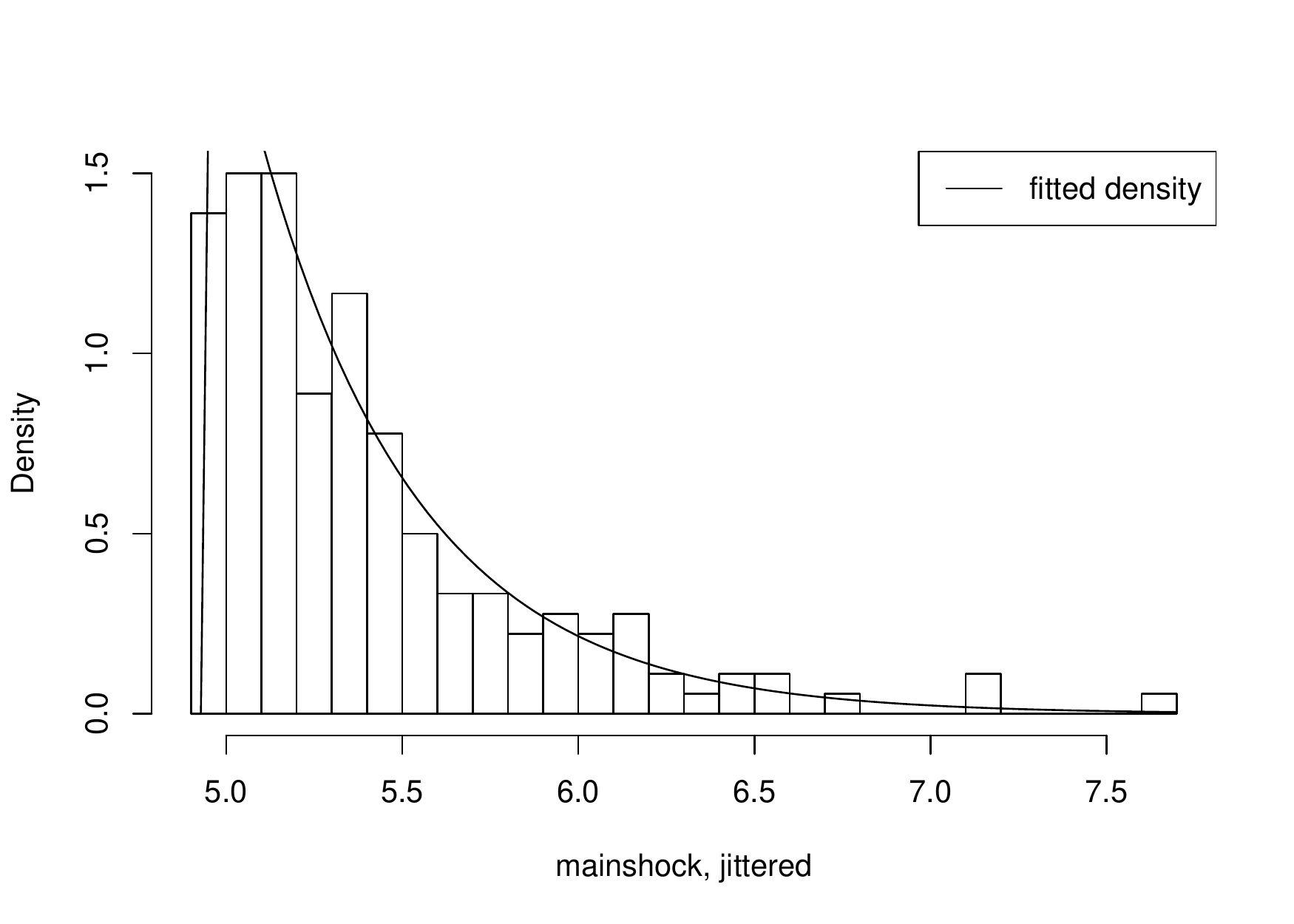}
\end{subfigure}
\begin{subfigure}{.49\textwidth}
\centering
\includegraphics[width=\linewidth]{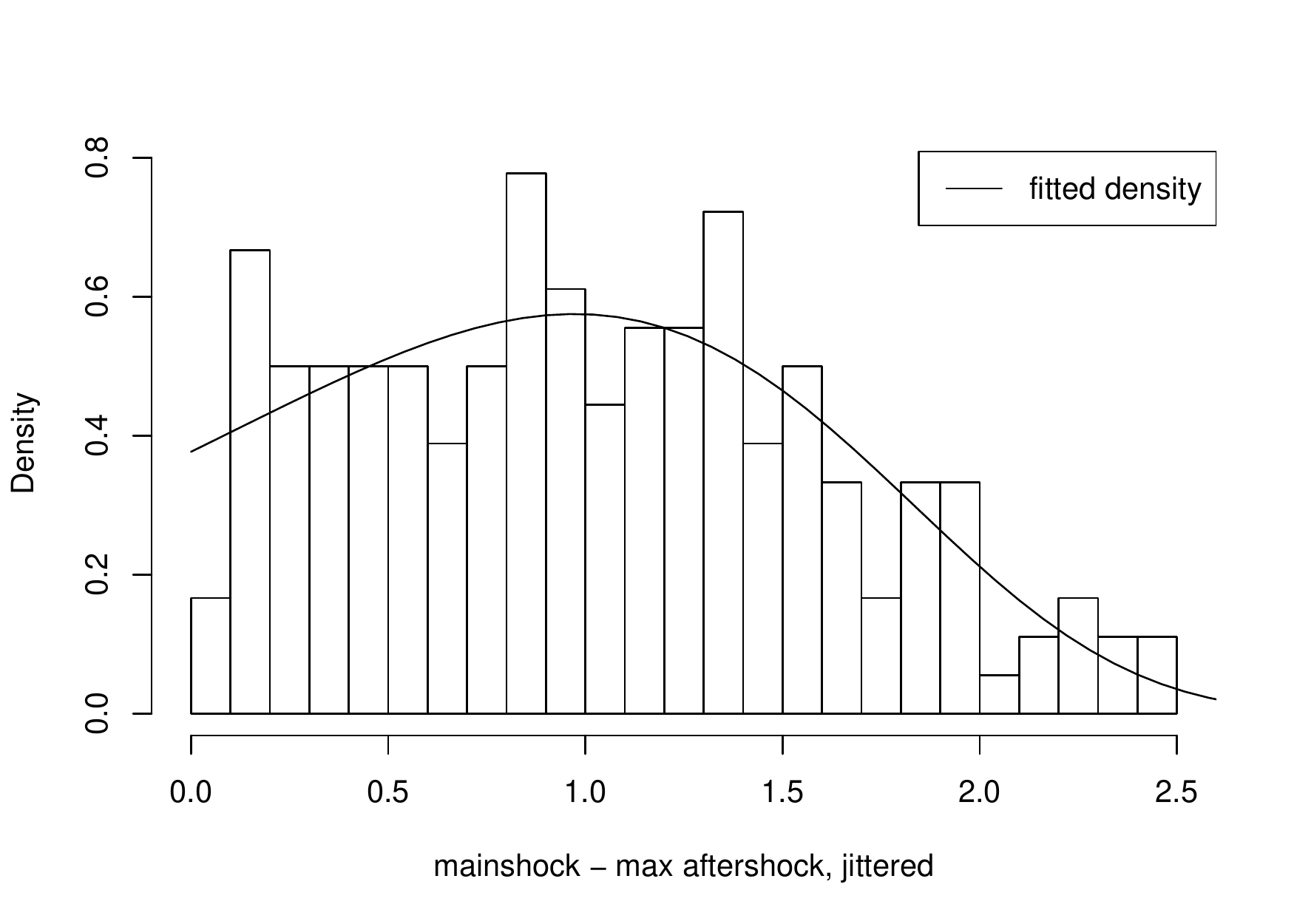}
\end{subfigure}
\caption{The histogram and fitted curve of: i) mainshocks with exponential distribution (left); ii) differences between mainshocks and largest aftershocks with Gompertz distribution (right).}
\label{fig:fit}
\end{figure}

The mainshock sequence $(x_i)$ is fitted with an exponential distribution truncated at 4.95 -- we take into consideration the continuity correction.  Since all observations are discrete by 0.1 increment, from now on whenever we show the fit of a distribution or calculate the goodness-of-fit $p$-value, we jitter all observations by uniform noises between $(-.05,.05)$.  The fit is shown on the left panel of Figure~\ref{fig:fit} and the Komogorov-Smirnov $p$-value is 0.83, indicating a good fit.

Next we fit a Gompertz distribution to the difference $x_i-y_i$ by maximizing the following censored likelihood:
$$
	L(\beta,C|x_i,y_i) = \prod_{y_i\ge4}f_Z(x_i-y_i;\beta,C) \prod_{y_i<4}(1-F_Z(x_i-4;\beta,C)),
$$
where $F_Z$ is as defined in \eqref{eq:gompertz} and $f_Z$ is the corresponding density.  To assess the goodness-of-fit, we first approximate the complete set of maximum aftershock sequence by $\tilde{y}_i$ as follows .  When $y_i\ge4$, set $\tilde{y}_i :=y_i$.  When $y_i<4$, simulate $z_i$ from $F$ conditional on $z_i \ge x_i - 4$ and set $\tilde{y}_i:= x_i - z_i$.  The histogram of jittered $\tilde{y}_i$ is shown on the right panel of Figure~\ref{fig:fit} with the fitted density.  The $p$-value is 0.95.

\begin{figure}[ht]
\centering
\includegraphics[width=.7\linewidth]{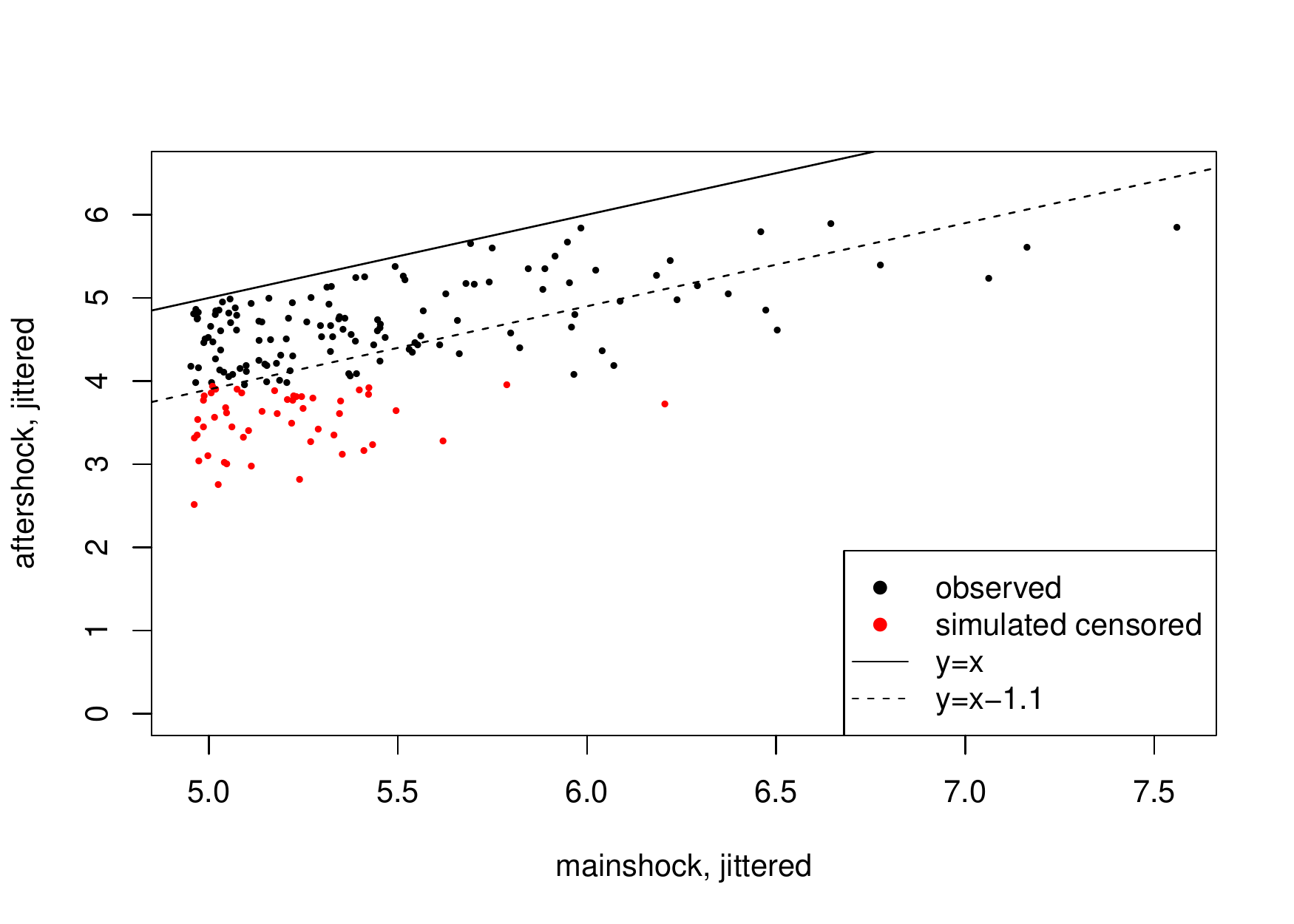}
\caption{Scatterplot of mainshocks vs.~aftershocks.  Black indicates observations and red indicates censored aftershocks with magnitude $<4$ and simulated from the fitted model.  The solid line indicate $y=x$.  Clearly all observations falls below as $Y\le X$ by definition.  The dotted line indicates $y=x-1.1$ as suggested by the B\r{a}th's law.}
\label{fig:fill}
\end{figure}

The scatterplot of the jittered $(x_i,\tilde{y}_i)$ is plotted in Figure~\ref{fig:fill}, with the red point indicating the simulation for the censored observations.  As we can see, the simulation is in agreement with the pattern of the observed pairs.  We also note that the B\r{a}th's law \cite{Bath1965} -- the empirical evidence that the magnitude difference between the mainshock and the largest aftershock is constant between 1.1 and 1.2 -- can be well-justified by the fitted model.  The fitted mean of $Y-X$ is 1.1. The $x=y+1.1$ line is shown in as the dotted line in Figure~\ref{fig:fill}.  

\subsection{Extreme value approach}

\begin{figure}[h]
\begin{center}
\includegraphics[width=0.5\linewidth]{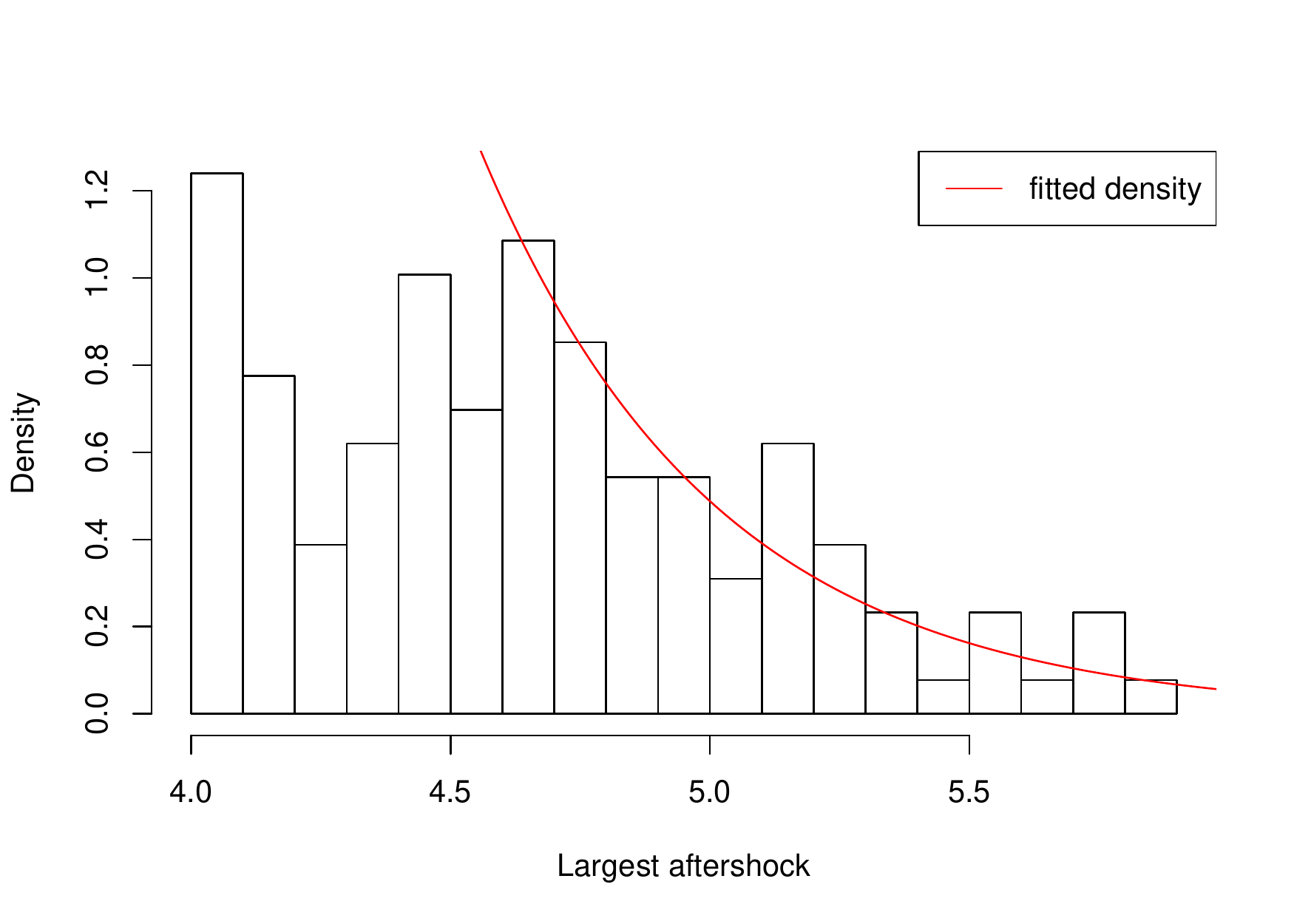}
\caption{The histogram of the largest aftershocks and fitted exponential density}
\label{fig:histY}
\end{center}
\end{figure}

For the extreme value analysis approach, we use the same estimate for the marginal distribution of $X$  as in the parametric approach.  We fit an exponential distribution truncated at 4.55 to the marginal distribution $Y$. The fitted density is shown in Figure \ref{fig:histY} with the Komogorov-Smirnov $p$-value 0.11. The fitted distributions are used to compute $p_1$ and $p_2$ in \eqref{eq:pst}.

When estimating $R(x, 1)$, we note that there are ties in the data as they are rounded to one decimal place.  For this, we randomly assign ranks to the tied observations. The missing values of $Y$ (they are censored above 4) does not effect the estimator in \eqref{eq: hatR} provided that $k$ is smaller than $n_1$, the number of observed $Y_i$'s. Obviously, the ranks of the missing values are smaller than $n-n_1$, thus the corresponding indicator function in \eqref{eq: hatR} equals to zero regardless of the precise value of $R_i^Y$. 

\begin{figure}[h]
\begin{center}
\includegraphics[width=0.35 \paperwidth]{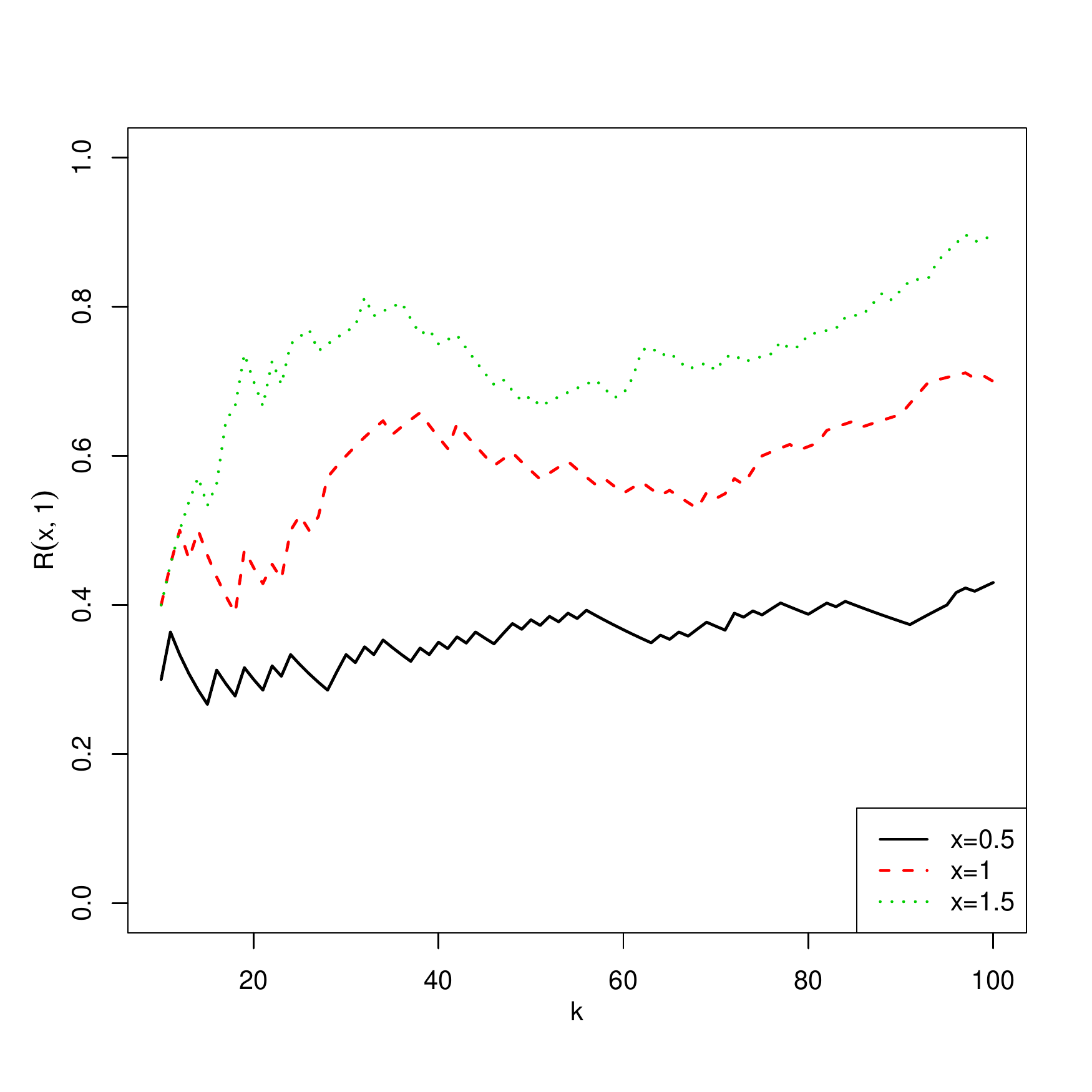}
\includegraphics[width=0.35 \paperwidth]{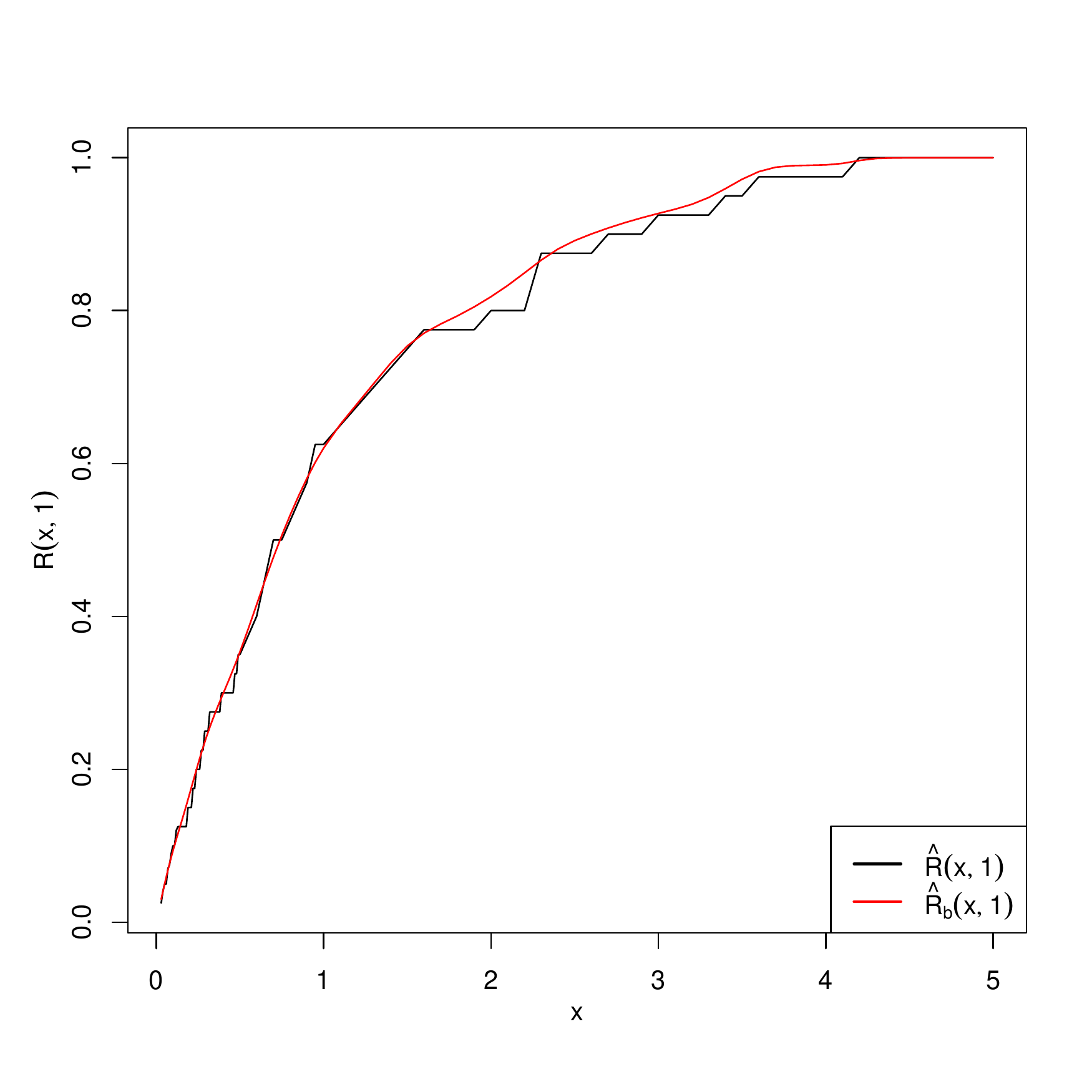}
\caption{Left: The estimator of $R(x, 1)$ is given in \eqref{eq: hatR}. Right: The estimator of $R(x, 1)$ is given in \eqref{eq: hatR}.}
\label{Fig:Rk}
\end{center}
\end{figure}

The left panel of Figure \ref{Fig:Rk} shows the estimates of $R(x,1)$ for three different values of $x$ and $k\in [10, 100]$. Also note that $R(1,1)$ is  a commonly used quantity to distinguish tail dependence ($R(1,1)>0$) and tail independence ($R(1,1)=0$). Roughly, tail dependence
says that the extremes of X and Y tend to occur simultaneously while joint extremes
rarely occur under tail independence.
This plot clearly suggests tail dependence between $X$ and $Y$ because the estimates of $R(1,1)$ are clearly above zero. Based on these three curves, we choose 
our $k=40$. 

With the choice of $k=40$, we obtain the non-parametric estimate of $R(x, 1)$ for $x\in[0.02, 5]$ plotted in the black curve in right panel of Figure \ref{Fig:Rk}.  The wiggly behaviour of this estimator motivates us to consider a smoothing method. We adopt the smoothing method introduced in \cite{Kiriliouketal2018}, which makes use of the beta copula.  This smoothed estimator, denoted as $\hat R_b(x,1)$, respects the pointwise upper bounds of the function, that is $R(x, 1)\leq \max (x, 1)$ and it does not require smoothing parameter such as bandwidth. 
 The resulted estimates are represented by the red curve in right panel of Figure \ref{Fig:Rk}. The two estimators are coherent with each other.  $\hat R_b(x,1)$ is only used in obtaining the level curves in Figure~\ref{fig:level}.

\subsection{Results and comparison}

We are ready to estimate probabilities for the joint tail of $(X,Y)$.  First, we estimate the tail probability\footnote{We remark that as our data set  consists of only mainshocks with magnitude ($X\geq 5$), the probability in this section has to be interpreted as a conditional probability that given a significant mainshock $X\ge5$ occurs. } defined in \eqref{eq:pst} for the ten largest earthquakes (mainshocks) in the NAFZ since 1965.  As shown in the fifth and sixth column of Table~\ref{tab:prob}, the estimates by two approaches are surprisingly close to each other, which supports the reasonability of the results. We emphasize that the two approaches only share one common assumption, that is, the marginal distribution of $X$.  The distribution of $Y$ and the dependence between the $(X, Y)$ are modelled separately. 

\begin{figure}[ht]
\centering
\includegraphics[width=.8\linewidth]{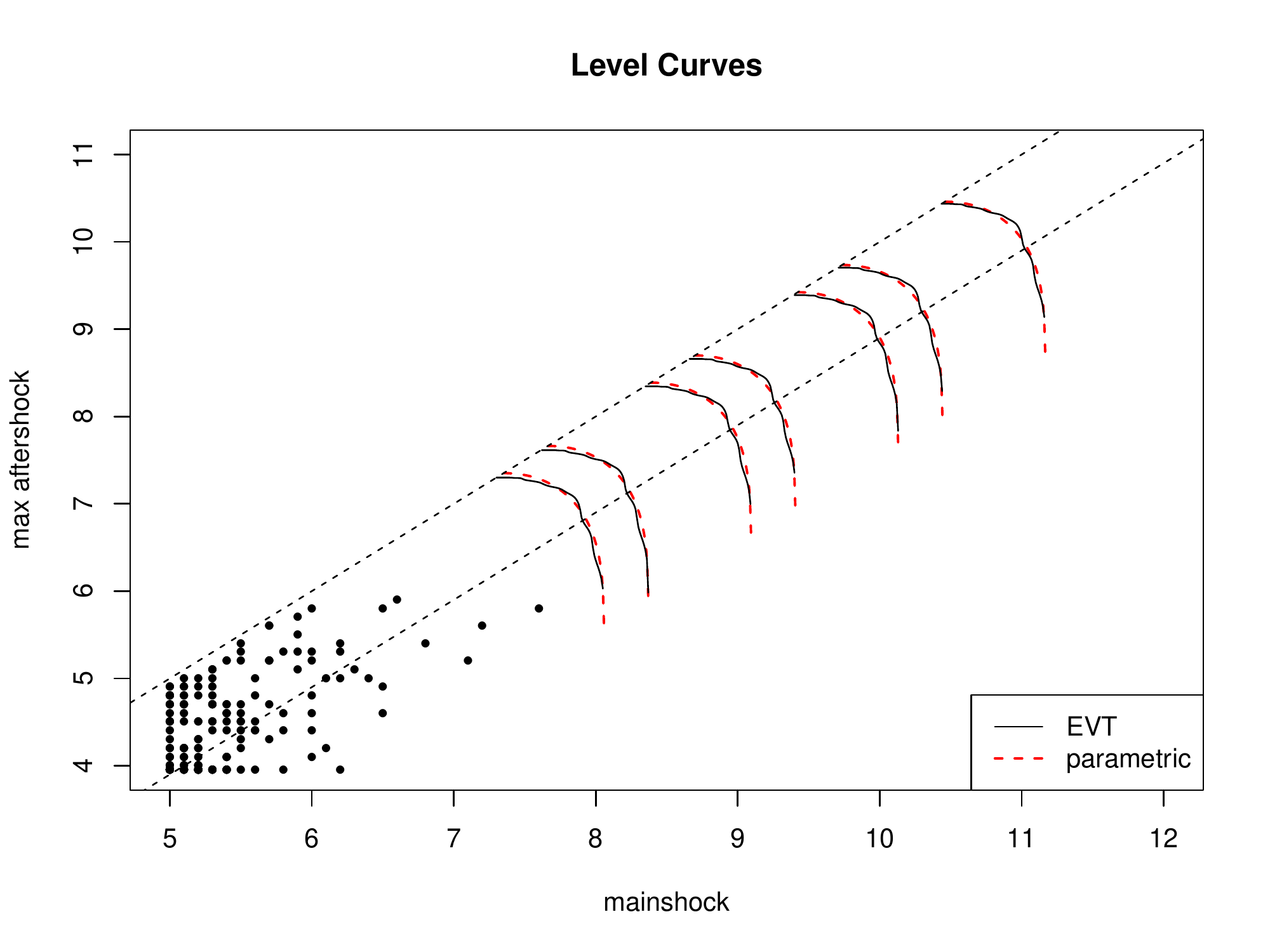}
\caption{Predicted level curve for $(X,Y)$ based on the parametric model (red dotted) and extreme value analysis (black solid), with existing observations.}
\label{fig:level}
\end{figure}

Next we obtain the level curves of $(X, Y)$ for the tail probability sequence 
$$(10^{-3},5\cdot10^{-4},10^{-4},5\cdot10^{-5},10^{-5},5\cdot10^{-6},10^{-6}),$$
as shown in Figure~\ref{fig:level}.
The points $(x, y)$ on each curve represents such that $\P(X>x,Y>y) = p$ for the given probability level.
 Albeit based on different theories, the two approaches provide coinciding prediction results.  The two dot lines in Figure ~\ref{fig:level} correspond to $x=y$ and $x=y+1.1$. The horizontal shape of the curves between these two lines indicates that the results respect the B\r{a}th's law.  This is particularly remarkable for the non-parametric approach, which does not impose any dependence structure for $(X, Y)$.

\begin{table}[ht]
\centering
\begin{tabular}{ccccccl}
  \hline
 &  & & largest & parametric & non-parametric &\\ 
 & date & mainshock  &  aftershock  & probability & probability & location \\
  \hline
1 & 1999-08-17 & 7.6 & 5.8 & 0.00265 & 0.00257 & \.{I}zmit \\ 
  2 & 1970-03-28 & 7.2 & 5.6 & 0.00618 & 0.00580 & Gediz \\ 
  3 & 1999-11-12 & 7.1 & 5.2 & 0.00815 & 0.00805 & D\"{u}zce \\ 
  4 & 1967-07-22 & 6.8 & 5.4 & 0.01413 & 0.01356 & Mudurnu \\ 
  5 & 1992-03-13 & 6.6 & 5.9 & 0.01429 & 0.01464 & Erzincan \\ 
  6 & 2002-02-03 & 6.5 & 5.8 & 0.01785 & 0.01827 & Afyon \\ 
  7 & 1969-03-28 & 6.5 & 4.9 & 0.02927 & 0.02745 & Ala\c{s}ehir \\ 
  8 & 1968-09-03 & 6.5 & 4.6 & 0.03092 & 0.03051 & Bartin \\ 
  9 & 1995-10-01 & 6.4 & 5.0 & 0.03437 & 0.03296 & Dinar \\ 
  10 & 2017-07-20 & 6.3 & 5.1 & 0.03938 & 0.03747 & Mugla Province \\ 
   \hline
\end{tabular}
\caption{Tail probability estimation for the ten largest earthquakes in the NAFZ since 1965.}
\label{tab:prob}
\end{table}


\section{Discussion}\label{Sec:Discuss}

In this paper we consider estimating the tail probability of an extreme earthquake event where the mainshock magnitude $X$ and the largest aftershock magnitude $Y$ both exceed certain thresholds.  We approach the problems from two directions.  On one hand, based on the well-known stochastic rules for aftershocks, we propose a joint parametric model for $(X,Y)$, estimate the model using (censored) maximum likelihood, and from the model, calculate the desired probabilities.  On the other hand, we use non-parametric methods from bivariate extreme value analysis to extrapolate tail probabilities.  We illustrates both methods using the earthquake data in the North Anatolian Fault Zone (NAFZ) in Turkey from 1965 to 2018.  The two approaches produce surprisingly agreeing results.  

This is an exploratory effort in applying multivariate extreme value analysis to seismology problems and much extension is possible.  For example, the occurrences of the earthquake events can be modelled in time and return level of extreme events can be estimated.  Further information, such as distance between shocks and other geological covariates, can be incorporated into the analysis to provide more accurate or customized results.  

This paper serves as a confirmation that simple techniques from multivariate extreme value analysis, though with little expert knowledge behind the data, is able to provide useful information in the analysis of extreme events. 

 
\section*{Acknowledgements}

The authors would like to thank Anna Kiriliouk for helpful discussion on a smoothed estimator of the stable tail dependence function.

\bibliographystyle{plainnat}
\bibliography{earthquake}

\begin{thebibliography}{33}
\providecommand{\natexlab}[1]{#1}
\providecommand{\url}[1]{\texttt{#1}}
\expandafter\ifx\csname urlstyle\endcsname\relax
  \providecommand{\doi}[1]{doi: #1}\else
  \providecommand{\doi}{doi: \begingroup \urlstyle{rm}\Url}\fi

\bibitem[Ambraseys(1970)]{Ambraseys1970}
N.~N. Ambraseys.
\newblock Some characteristic features of the {Anatolian} fault zone.
\newblock \emph{Tectonophysics}, 9\penalty0 (2-3):\penalty0 143--165, 1970.

\bibitem[Ambraseys and Finkel(1987)]{AmbraseysFinkel1987}
N.N. Ambraseys and C.F. Finkel.
\newblock {Seismicity of Turkey and neighbouring regions, 1899-1915}.
\newblock In \emph{Annales geophysicae. Series B. Terrestrial and planetary
  physics}, volume~5, pages 701--725, 1987.

\bibitem[B{\aa}th(1965)]{Bath1965}
M.~B{\aa}th.
\newblock Lateral inhomogeneities of the upper mantle.
\newblock \emph{Tectonophysics}, 2\penalty0 (6):\penalty0 483--514, 1965.

\bibitem[Beirlant et~al.(2004)Beirlant, Goegebeur, Segers, and
  Teugels]{Beirlantetal2004}
J.~Beirlant, Y.~Goegebeur, J.~Segers, and J.~Teugels.
\newblock \emph{Statistics of Extremes, Theory and Applications}.
\newblock Chichester: Wiley, 2004.

\bibitem[Beirlant et~al.(2016)Beirlant, Escobar-Bach, Goegebeur, and
  Guillou]{Beirlantetal2016}
J.~Beirlant, M.~Escobar-Bach, Y.~Goegebeur, and A.~Guillou.
\newblock Bias-corrected estimation of stable tail dependence function.
\newblock \emph{Journal of Multivariate Analysis}, 143:\penalty0 453--466,
  2016.

\bibitem[Beirlant et~al.(2018)Beirlant, Kijko, Reynkens, and
  Einmahl]{Beirlantetal2018}
J.~Beirlant, A.~Kijko, T.~Reynkens, and J.H.J. Einmahl.
\newblock {Estimating the maximum possible earthquake magnitude using extreme
  value methodology: the Groningen case}.
\newblock \emph{Natural Hazards}, pages 1--23, 2018.

\bibitem[B{\"u}cher et~al.(2011)B{\"u}cher, Dette, and
  Volgushev]{Bucheretal2011}
A.~B{\"u}cher, H.~Dette, and S.~Volgushev.
\newblock New estimators of the pickands dependence function and a test for
  extreme-value dependence.
\newblock \emph{The Annals of Statistics}, 39\penalty0 (4):\penalty0
  1963--2006, 2011.

\bibitem[Cap{\'e}ra{\`a} et~al.(1997)Cap{\'e}ra{\`a}, Foug{\`e}res, and
  Genest]{Caperaaetal1997}
P.~Cap{\'e}ra{\`a}, A.-L. Foug{\`e}res, and C.~Genest.
\newblock A nonparametric estimation procedure for bivariate extreme value
  copulas.
\newblock \emph{Biometrika}, 84\penalty0 (3):\penalty0 567--577, 1997.

\bibitem[Cui et~al.(2011)Cui, Chen, Zhu, Su, Wei, Han, Liu, and
  Zhuang]{cui2011}
P.~Cui, X.-Q. Chen, Y.-Y. Zhu, F.-H. Su, F.-Q. Wei, Y.-S. Han, H.-J. Liu, and
  J.-Q. Zhuang.
\newblock {The Wenchuan earthquake (May 12, 2008), Sichuan province, China, and
  resulting geohazards}.
\newblock \emph{Natural Hazards}, 56\penalty0 (1):\penalty0 19--36, 2011.

\bibitem[de~Haan and de~Ronde(1998)]{dehaan1998}
L.~de~Haan and J.~de~Ronde.
\newblock Sea and wind: multivariate extremes at work.
\newblock \emph{Extremes}, 1\penalty0 (1):\penalty0 7, 1998.

\bibitem[de~Haan and Ferreira(2006)]{deHaanFerreira2006}
L.~de~Haan and A.~Ferreira.
\newblock \emph{Extreme Value Theory: An Introduction}.
\newblock Berlin: Springer, 2006.

\bibitem[de~Haan and Resnick(1977)]{deHaanResnick1977}
L.~de~Haan and S.~I. Resnick.
\newblock Limit theory for multivariate sample extremes.
\newblock \emph{Zeitschrift f{\"u}r Wahrscheinlichkeitstheorie und verwandte
  Gebiete}, 40\penalty0 (4):\penalty0 317--337, 1977.

\bibitem[Einmahl et~al.(2008)Einmahl, Krajina, and Segers]{Einmahletal2008}
J.H.J. Einmahl, A.~Krajina, and J.~Segers.
\newblock A method of moments estimator of tail dependence.
\newblock \emph{Bernoulli}, 14\penalty0 (4):\penalty0 1003--1026, 2008.

\bibitem[Foug{\`e}res et~al.(2015)Foug{\`e}res, De~Haan, and
  Mercadier]{Fougeresetal2015}
A.-L. Foug{\`e}res, L.~De~Haan, and C.~Mercadier.
\newblock Bias correction in multivariate extremes.
\newblock \emph{The Annals of Statistics}, 43\penalty0 (2):\penalty0 903--934,
  2015.

\bibitem[Gardner and Knopoff(1974)]{gardner1974}
J.K. Gardner and L.~Knopoff.
\newblock {Is the sequence of earthquakes in Southern California, with
  aftershocks removed, Poissonian?}
\newblock \emph{Bulletin of the Seismological Society of America}, 64\penalty0
  (5):\penalty0 1363--1367, 1974.

\bibitem[Gutenberg and Richter(1944)]{GutenbergRichter1944}
B.~Gutenberg and C.~F. Richter.
\newblock Frequency of earthquakes in {California}.
\newblock \emph{Bulletin of the Seismological Society of America}, 34\penalty0
  (4):\penalty0 185--188, 1944.

\bibitem[Hanks and Kanamori(1979)]{hanks1979}
T.~C. Hanks and H.~Kanamori.
\newblock A moment magnitude scale.
\newblock \emph{Journal of Geophysical Research: Solid Earth}, 84\penalty0
  (B5):\penalty0 2348--2350, 1979.

\bibitem[Kanamori(1977)]{kanamori1977}
H.~Kanamori.
\newblock The energy release in great earthquakes.
\newblock \emph{Journal of geophysical research}, 82\penalty0 (20):\penalty0
  2981--2987, 1977.

\bibitem[Kijko(2004)]{Kijko2004}
A.~Kijko.
\newblock Estimation of the maximum earthquake magnitude, m max.
\newblock \emph{Pure and Applied Geophysics}, 161\penalty0 (8):\penalty0
  1655--1681, 2004.

\bibitem[Kiriliouk et~al.(2018)Kiriliouk, Segers, and
  Tafakori]{Kiriliouketal2018}
A.~Kiriliouk, J.~Segers, and L.~Tafakori.
\newblock An estimator of the stable tail dependence function based on the
  empirical beta copula.
\newblock \emph{Extremes}, 21\penalty0 (4):\penalty0 581--600, 2018.

\bibitem[Ledford and Tawn(1997)]{ledford1997}
A.~W. Ledford and J.~A. Tawn.
\newblock Modelling dependence within joint tail regions.
\newblock \emph{Journal of the Royal Statistical Society: Series B (Statistical
  Methodology)}, 59\penalty0 (2):\penalty0 475--499, 1997.

\bibitem[Marza(2004)]{marza2004}
V.I. Marza.
\newblock {On the death toll of the 1999 Izmit (Turkey) major earthquake}.
\newblock \emph{ESC General Assembly Papers, European Seismological Commission,
  Potsdam}, 2004.

\bibitem[Ogata(1988)]{Ogata1988}
Y.~Ogata.
\newblock Statistical models for earthquake occurrences and residual analysis
  for point processes.
\newblock \emph{Journal of the American Statistical association}, 83\penalty0
  (401):\penalty0 9--27, 1988.

\bibitem[Parsons et~al.(2000)Parsons, Toda, Stein, Barka, and
  Dieterich]{Parsonsetal2000}
T.~Parsons, S.~Toda, R.~S. Stein, A.~Barka, and J.~H Dieterich.
\newblock Heightened odds of large earthquakes near istanbul: An
  interaction-based probability calculation.
\newblock \emph{Science}, 288\penalty0 (5466):\penalty0 661--665, 2000.

\bibitem[Polat et~al.(2002)Polat, Eyidogan, Haessler, Cisternas, and
  Philip]{Polatetal2002}
O.~Polat, H.~Eyidogan, H.~Haessler, A.~Cisternas, and H.~Philip.
\newblock Analysis and interpretation of the aftershock sequence of the {August
  17, 1999, Izmit (Turkey)} earthquake.
\newblock \emph{Journal of Seismology}, 6\penalty0 (3):\penalty0 287--306,
  2002.

\bibitem[Poon et~al.(2004)Poon, Rockinger, and Tawn]{PoonRockingerTawn2004}
S.-H. Poon, M.~Rockinger, and J.~Tawn.
\newblock Extreme-value dependence in financial markets: diagnostics, models
  and financial implications.
\newblock \emph{Review of Financial Studies}, 17:\penalty0 581 -- 610, 2004.

\bibitem[Reasenberg and Jones(1989)]{ReasenbergJones1989}
P.~A Reasenberg and L.~M. Jones.
\newblock Earthquake hazard after a mainshock in {California}.
\newblock \emph{Science}, 243\penalty0 (4895):\penalty0 1173--1176, 1989.

\bibitem[Reilinger et~al.(2000)Reilinger, Ergintav, B{\"u}rgmann, McClusky,
  Lenk, Barka, Gurkan, Hearn, Feigl, and Cakmak]{Reilingeretal2000}
R.E. Reilinger, S.~Ergintav, R.~B{\"u}rgmann, S.~McClusky, O.~Lenk, A.~Barka,
  O.~Gurkan, L.~Hearn, K.L. Feigl, and R.~Cakmak.
\newblock Coseismic and postseismic fault slip for the {17 August 1999, M= 7.5,
  Izmit, Turkey earthquake}.
\newblock \emph{Science}, 289\penalty0 (5484):\penalty0 1519--1524, 2000.

\bibitem[Stein et~al.(1997)Stein, Barka, and Dieterich]{Steinetal1997}
R.~S. Stein, A.~A Barka, and J.~H. Dieterich.
\newblock Progressive failure on the {North Anatolian} fault since 1939 by
  earthquake stress triggering.
\newblock \emph{Geophysical Journal International}, 128\penalty0 (3):\penalty0
  594--604, 1997.

\bibitem[Utsu(1970)]{Utsu1970}
T.~Utsu.
\newblock Aftershocks and earthquake statistics (1): Some parameters which
  characterize an aftershock sequence and their interrelations.
\newblock \emph{Journal of the Faculty of Science, Hokkaido University. Series
  7, Geophysics}, 3\penalty0 (3):\penalty0 129--195, 1970.

\bibitem[Utsu(1971)]{Utsu1971}
T.~Utsu.
\newblock Aftershocks and earthquake statistics (2): Further investigation of
  aftershocks and other earthquake sequences based on a new classification of
  earthquake sequences.
\newblock \emph{Journal of the Faculty of Science, Hokkaido University. Series
  7, Geophysics}, 3\penalty0 (4):\penalty0 197--266, 1971.

\bibitem[Utsu(1972)]{Utsu1972}
T.~Utsu.
\newblock Aftershocks and earthquake statistics (3): Analyses of the
  distribution of earthquakes in magnitude, time and space with special
  consideration to clustering characteristics of earthquake occurrence (1).
\newblock \emph{Journal of the Faculty of Science, Hokkaido University. Series
  7, Geophysics}, 3\penalty0 (5):\penalty0 379--441, 1972.

\bibitem[Vere-Jones et~al.(2006)Vere-Jones, Murakami, and
  Christophersen]{Vere2006}
D.~Vere-Jones, J.~Murakami, and A.~Christophersen.
\newblock {A further note on Bath’s law}.
\newblock In \emph{The 4th International Workshop on Statistical Seismology
  (Statsei4)}, pages 611--0011, 2006.

\end{thebibliography}
\end{document}